\documentclass[doublecol]{epl2}

\usepackage{subfigure}
\usepackage{epic}\usepackage{eepic}
\usepackage{bm}
\usepackage[dvips]{epsfig}
\usepackage{graphicx} \usepackage{amsmath} \usepackage{amssymb}
\usepackage{abraces}

\usepackage{dcolumn}

\newcommand{\comment}[1]{}
\newcommand{\lp}{\underline{p}}
\newcommand{\lo}{\underline{\omega}}
\newcommand{\up}{\overline{p}}
\newcommand{\uo}{\overline{\omega}}

\newcommand{\be}{\begin{eqnarray}}
\newcommand{\ee}{\end{eqnarray}}

\newcommand{\ba}{\begin{align}}
\newcommand{\ea}{\end{align}}

\newcommand{\bg}{\begin{gather}}
\newcommand{\eg}{\end{gather}}

\newcommand{\nn}{\nonumber \\}

\newcommand{\no}{\label}

\renewcommand{\P}{\hat{P}}
\newcommand{\Q}{\hat{Q}}

\newcommand{\p}{\widetilde{P}}
\newcommand{\q}{\widetilde{Q}}

\newcommand{\ot}{\otimes}
\newcommand{\oo}{\bullet}

\title{Quantum non-locality co-exists with locality}

\author{Armen E. Allahverdyan and Arshag Danageozian}
\institute{Yerevan Physics Institute, 2 Alikhanian Brothers street, Yerevan
0036, Armenia \footnote{To be published in EPL (Europhysics Letters) (2018).}} 
\shortauthor{Allahverdyan and Danageozian}

\abstract{ Quantum non-locality is normally defined via violations
of Bell's inequalities that exclude certain classical hidden variable
theories from explaining quantum correlations. Another definition of
non-locality refers to the wave-function collapse thereby one can prepare
a quantum state from arbitrary far away. In both cases one can debate on
whether non-locality is a real physical phenomenon, e.g. one can
employ formulations of quantum mechanics that does not use collapse, or
one can simply refrain from explaining quantum correlations via
classical hidden variables. Here we point out that there is a non-local
effect within quantum mechanics, i.e. without involving hidden variables
or collapse. This effect is seen via imprecise (i.e. interval-valued)
joint probability of two observables, which replaces the ill-defined
notion of the precise joint probability for non-commuting observables.
It is consistent with all requirements for the joint probability, e.g.
those for commuting observales.  The non-locality amounts to a fact
that (in a two-particle system) the joint imprecise probability of
non-commuting two-particle observables (i.e. tensor product of
single-particle observables) does not factorize into single-particle
contributions, even for uncorrelated states of the two-particle system.
The factorization is recovered for a less precise (i.e.  the one involving a
wider interval) joint probability. This approach to non-locality
reconciles it with locality, since the latter emerges as a less precise
description. }


\begin{document}

\maketitle

Physics holds signal locality thereby the state of an isolated system
(particle) cannot be changed by manipulating other particles that
interacted with it in the past. Attempts to make this notion stronger
created a bunch of quantum concepts known as non-locality
\cite{deMuynck,jarrett,peinaar}. Two main aspects of non-locality were
identified: {\it (i)} application of the projection postulate to an
entangled subsystem that changes (steers) the state of another subsystem
arbitrary far away \cite{deMuynck}. {\it (ii)} Bell's inequalities,
their derivation and interpretation \cite{deMuynck}. Despite of a huge
effort devoted to the issue of non-locality, there is no general
agreement on whether it really exists
\cite{brukner,vindi,grif,evo,bachus}. 

Indeed, {\it (i)} does depend on the specific interpretation adopted for
quantum mechanics, e.g. it is present in the Copenhagen interpretation,
but absent in the many-world interpretation \cite{evo,bachus} and in the
consistent histories interpretation \cite{grif}. Note that the issue of
collapse (i.e.  what happens to the quantum system after measurement) is
not a part of the quantum probability given by Born's rule. The latter
can and does apply for describing experiments without worrying about
what happens after measurement. This is one reason why the attention in
studying non-locality shifted to Bell's inequalities that do not assume
the collapse. Instead they focus on explaining quantum correlations via
hidden variables, i.e. they assume structures that are additional with
respect to quantum probability. Bell's inequalities can be derived
assuming that the hidden variables|in addition to signal locality|hold
the outcome independence feature (frequently also called local
causality) \cite{jarrett,peinaar,fine}.  Alternatively, Bell's
inequality can be derived without hidden variables, but assuming that
frequencies obtained via different non-commuting measurements can be
embedded into a single Kolmogorovian space \cite{eberhard}. Another
version of this condition is that several (more than two) non-commuting
quantities do have a joint probability \cite{fine,brukner}.  The
joint-probability assumptions can be imposed also on a single quantum
particle, where no space-separated sub-systems exist. This leads to
single-particle Bell inequalities \cite{gudder_bell,accardi}. Then,
experimental validation of the Bell inequalities show that those
additional concepts do not apply to quantum mechanics. Is this the only
meaning of non-locality that certain interpretations have a cost to pay,
or that specific classic concepts do not apply? 

\comment{
The notion of contextuality to some extent repeated the historical
development of non-locality \cite{kochen,spe,hakop}. Consider Born's
probability ${\rm tr}(\rho P)$ for the probability of the eigenvalue $1$
of a projector $P$ in a quantum state with a density matrix $\rho$.
This probability is non-contextual, because it does not depend on an
observable (self-adjoint operator) whose eigen-projector is $P$.  The
choice of such an observable is not unique in Hilbert spaces with
dimension larger than 2.  Moreover, different observables having the
same eigen-projector need not commute.  Attempts to explain quantum
mechanics via hidden-variable theories created the notion of
contextuality, since it appeared that deterministic hidden-variables
explaining quantum probability cannot be non-contextual for Hilbert
spaces with ${\rm dim}>2$ \cite{kochen}. Technically, this
Kochen-Specker theorem amounts to prescribing to projectors values $0$
or $1$, respecting the sums of commuting projectors, and then running
into a contradiction.  The notion of contextuality was generalized
several times (see \cite{spe,hakop} for recent reviews) referring to
more general hidden-variable theories. Kochen-Specker theorem and Bell's
inequality are related, e.g. both exclude joint probabilities for several (more
than two) non-commuting observables \cite{fine}. Since the standard
contextuality refers to structures beyond quantum mechanics (i.e.
hidden-variable theories), its implications strictly within
quantum mechanics is an open question. }

We aim to demonstrate non-locality within quantum mechanics without
referring to hidden variables, imposed Kolmogorovian probability space
or projection postulate (hence it does not depend on which
interpretation one prescribes to).  The core of this demonstration is
that while two non-commuting projectors do not have a joint {\it
precise} (i.e. usual) probability within quantum mechanics, they do have
an imprecise (i.e.  interval-valued) joint probability
\cite{armen,arshag}. Then non-locality is seen in the fact that the
imprecise joint probability operator calculated for tensor products of
single-particle (local) observables does not reduce to a tensor product
of single-particle operators. Hence the imprecise joint probability
(calculated via the ordinary Born's rule) does not reduce to a product
of single-particle contributions, even if the state of two-particle
system is not correlated. Our results achieve on the level of imprecise
joint statistics what the previous results could not do on the level of
precise joint probability for non-commuting observables, since the
latter does not exist. This approach reconciles non-locality with
locality, because the local (tensor-product based) imprecise joint
probability appears to predict a wider interval: it is not wrong (as
compared to the non-local one), it is just less precise. 
We continue by recalling pertinent features of projectors (see
\cite{adler} for an accessible review), and then the notion of quantum
imprecise probability.

{\bf Projectors} are self-adjoint operators $P$ with $P^2=P$. Any projector
$P$ in a Hilbert space ${\cal H}$ bijectively relates to the sub-space
${\cal S}_P$ of ${\cal H}$ \cite{jauch_book}: 
\be
\no{bu}
{\cal S}_P=\{ |\psi\rangle \in {\cal H}; P|\psi\rangle=|\psi\rangle \}.
\ee
Eigenvalues of $P$ are $0$ and/or $1$, and it is a quantum analogue of the
characteristic function for a classical set \cite{jauch_book}. Hence
projectors define quantum probability: with a density
matrix $\rho$, the probability of finding the eigenvalue $1$
of $P$ is given by Born's formula ${\rm tr}(\rho P)$.

The simplest projectors are $0$ and $I$. For two projectors $P$ and $P'$ we define
\be
P\geq P'~~{\rm means}~~\langle \psi|P-P'|\psi\rangle\geq 0~~{\rm for~any}~~|\psi\rangle.
\no{grum}
\ee
Now apply (\ref{grum}) with $|\psi\rangle=|\psi_0\rangle$, where $P|\psi_0\rangle=0$,
and then with $|\psi\rangle=|\psi_1\rangle$, where $P'|\psi_1\rangle=|\psi_1\rangle$.
Hence the eigenvalues of $P$ and $P'$ relate to each other leading to
\be
\no{sar}
PP'=P'P=P'~~{\rm if}~~ P\geq P'.
\ee
Projectors (generally non-commuting) support logical operations
\cite{jauch_book}: negation $P^\perp=I-P$, conjunction $P\land Q$
(${\cal S}_{P\land Q}={\cal S}_{Q}\cap {\cal S}_{P}$ contains only those
vectors that belong both to ${\cal S}_{P}$ and ${\cal S}_{Q}$), and
disjunction $P\lor Q$, where ${\cal S}_{P\lor Q}$ contains all linear
combinations of vectors from ${\cal S}_{P}$ and from ${\cal S}_{Q}$.
There are alternative representations \cite{jauch_book}:
\begin{align}
  \label{bars}
  P\land Q={\rm max}_R[\, R\,|\,R^2=R, \, R\leq P, \, R\leq Q\,], \\
  P\lor Q={\rm  min}_R[\, R\,|\,R^2=R, \, R\geq P, \, R\geq Q\,],
  \label{colibri}
\end{align}
where the maximization and minimization go over projectors $R$
\cite{jauch_book}.  Indeed, if $R\leq P$, and $R\leq Q$ in (\ref{bars}),
then due to (\ref{bu},\ref{sar}), ${\cal S}_R$ is a subspace of both
${\cal S}_P$ and ${\cal S}_Q$. The maximal such subspace is ${\cal
S}_{P\land Q}$. Likewise, if $R\geq P$, and $R\geq Q$, then ${\cal S}_R$
has to contain both ${\cal S}_P$ and ${\cal S}_Q$. The smallest such
subspace is ${\cal S}_{P\lor Q}$, because joining ${\cal S}_{P}$ and
${\cal S}_{Q}$ as sets does not result to a linear space. 

Now $P\lor Q=0$ only if $P=Q=0$, but $P\land Q$ can be zero also for
non-zero $P$ and $Q$; e.g. in ${\rm dim}{\cal
H}=2$, we have either $P=Q$ or $P\land Q=0$ (and then $P\lor Q=I$). 

The above three operations are related with each other and with a limiting 
process \cite{jauch_book}:
\be
\no{gogi}
P\lor Q= (P^\perp\land Q^\perp)^\perp, \\
P\land Q={\rm lim}_{n\to\infty}(PQ)^n.
\no{nogi}
\ee
For $[P,Q]\equiv PQ-QP=0$ we have from (\ref{sar}--\ref{colibri}) 
the ordinary features of classical characteristic functions
\be
P\land Q=PQ,\quad P\lor Q=P+Q-PQ.
\no{dum}
\ee

{\bf Imprecise classical probability} generalizes the usual (precise)
probabilities \cite{bumbarash,intro}: the measure of uncertainty for an
event $E$ is an interval $[\lp(E),\up(E)]$, where
$0\leq\lp(E)\leq \up(E)$ are called lower and upper probabilities,
respectively. Now $\lp(E)$ (resp. $1-\up(E)$) is a measure of a sure
evidence in favor (resp. against) of $E$. The event $E$ is surely more
probable than $E'$, if $\lp(E)\geq \up(E')$. The usual probability is
recovered for $\lp(E)=\up(E)$. Two different pairs $[\lp(E),\up(E)]$
and $[\lp'(E),\up'(E)]$ can hold simultaneously (i.e they are {\it
  consistent}) if
\begin{eqnarray}
  \label{eq:100}
\lp'(E)\leq \lp(E)~~{\rm and}~~
\up'(E)\geq \up(E).  
\end{eqnarray}
Every probability is consistent with $\lp'(E)=0$, $\up'(E)=1$.
This non-informative situation is not described by the usual
theory \cite{bumbarash} that inadequately offers for it 
the homogeneous probability \cite{bumbarash}. Now $[\lp(E),\up(E)]$ does
not imply that there is an explicit (but possibly unknown) precise
probability for $E$ located in between $\lp(E)$ and $\up(E)$
\cite{papa}. 

{There are various imprecise classical probability models
\cite{intro,huber,suppes_zanotti_erken,dempster,shafer}.  They do have
numerous applications e.g. in decision making and artificial
intelligence \cite{intro}. Some of them were applied phenomenologically
for describing aspects of the Bell's inequality
\cite{suppes_zanotti,vourdas,fletcher}. The quantum imprecise
probability is to be sought independently, along the physical arguments.}
Below we recall how it is determined. 

{\bf Imprecise quatum probability}. For two non-commuting projectors $P$
and $Q$ one looks for upper $\uo(P,Q)$ and lower $\lo(P,Q)$ non-negative
probability operators. For a state with density matrix $\rho$, the
respective upper and lower probabilities are given by Born's rule:
\begin{eqnarray}
  \label{borno}
\up(P,Q)={\rm
  tr}(\rho\,\uo(P,Q)), \quad \lp(P,Q)={\rm tr}(\rho\,\lo(P,Q)).  
\end{eqnarray}
The linearity of (\ref{borno}) over $\rho$ means that within the standard measurement
theory $\up(P,Q)$ ($\lp(P,Q)$) can be determined via measuring Hermitean operator
$\uo(P,Q)$ ($\lo(P,Q)$) in a state with an unknown $\rho$ \cite{deMuynck}.
The following requirements determine $\uo(P,Q)$ and $\lo(P,Q)$
\cite{armen} 
\begin{align}
  \label{eq:1}
&  0\leq \lo(P, Q)=\lo(Q, P)\leq \uo(P, Q)=\uo(Q, P)\leq
I. \\ \nonumber\\
  \label{eq:2}
&  [\,\omega(P, Q),Q\,]=[\,\omega(P, Q),P\,]=0 ~{\rm for}~ \omega=\lo,\uo.\\ 
\nonumber\\  
  \label{eq:3}
&  \lo(P, Q)= \uo(P, Q)=P Q~~~ {\rm if}~~~ [P, Q]=0. \\ \nonumber\\
  \label{eq:4}
& {\rm tr}(\rho \,\lo(P, Q)\,)\leq {\rm tr}(\rho\, P
Q)\leq{\rm tr}(\rho\,\uo(P, Q)\,) \\ 
&{\rm if}~~ [P_a, \rho]=0 ~~{\rm or}~~ [Q_b,\rho]=0.
  \label{eq:5}
\end{align}
Eqs.~(\ref{eq:1}, \ref{borno}) forces $0\leq\lp(P,Q)\leq \up(P,Q)\leq 1$
for any $\rho$. Eq.~(\ref{eq:1}) also demands symmetry
with respect to $P$ and $Q$ that is necessary for the joint probability.  

Let us recall the non-contextuality feature of quantum probability: the
ordinary Born's probability ${\rm tr}(\rho P)$ is non-contextual,
because it does not depend on an observable (Hermitean operator) $A$
whose eigen-projector is $P$.  The choice of $A$ is not unique in
Hilbert spaces with dimension larger than 2 (different observables
having the same eigen-projector need not commute). Now $\lo(P, Q)$ and
$\uo(P, Q)$ are non-contextual in the sense that they depend only on $P$
and $Q$; see also (\ref{borno}). A stronger feature holds:
(\ref{eq:2}) shows that both $\lo(P, Q)$ and $\uo(P, Q)$ can be measured
together with either $P$ or $Q$; cf.~(\ref{eq:18}). 

For $[P,Q]=0$ we revert to the usual joint probability; see
(\ref{eq:3}). For $Q=I$ we get from (\ref{eq:3}, \ref{borno}) the
marginal and precise probability of $P$.  Eqs.~(\ref{eq:4}, \ref{eq:5})
also refer to the consistency [in the sense of (\ref{eq:100})] with the
precise probability, because the latter is well-defined not only for
$[P,Q]=0$, but also under conditions (\ref{eq:5}), where it amounts to
${\rm tr}(\rho P Q)$. The latter can be calculated as an average 
${\rm tr}(\rho \frac{PQ+QP}{2})$ of the Hermitean operator 
$\frac{P Q+QP}{2}$.

Eqs.~(\ref{eq:1}--\ref{eq:4}) suffice for deducing \cite{armen}:
\begin{eqnarray}
  \label{eq:20}
  \lo(P, Q)=P\land Q, \qquad   \uo(P, Q)=P\lor Q-(P-Q)^2,~
\end{eqnarray}
where (\ref{eq:20}) are (resp.) the largest and the smallest positive
operators holding (\ref{eq:1}--\ref{eq:4}). Now $\lo(P, Q)$ is a
projector, while $\uo(P, Q)$ is generally just a non-negative operator.
For $[P,Q]=0$ we deduce (\ref{eq:3}) from (\ref{dum}) and (\ref{eq:20}). 

Eqs.~(\ref{eq:20}) imply (\ref{eq:2}), because|as follows from (\ref{bars}, 
\ref{colibri}) and checked directly|$P\land Q$, $P\lor Q$ and
$(P-Q)^2$ commute with each other and with $P$ and $Q$. Hence
\begin{eqnarray}
  \label{eq:18}
  [\,\uo(P, Q),\,\lo(P, Q)\,]=0. 
\end{eqnarray}
The origin of (\ref{eq:20}) is understood from (\ref{eq:1}, \ref{eq:2})
and (\ref{bars}, \ref{colibri}), i.e. $P\land Q$ and $P\lor Q$ qualify
as certain (resp.) lower and upper probability operators, while the
factor $(P-Q)^2$ in (\ref{eq:20}) is needed to ensure (\ref{eq:3}). 

An important geometric feature of (\ref{eq:20}) is that both
$PQP$ (i.e. the restriction of $Q$ into ${\cal S}_P$) and 
$QPQ$ hold:
\begin{eqnarray}
  \label{eq:22}
  \lo(P,Q)\leq PQP,\, QPQ\,\leq   \uo(P,Q).
\end{eqnarray}
Now $\lo(P,Q)\leq PQP$ is shown from $P\land Q\leq Q$ [see (\ref{bars})]
that implies $P\land Q=P(P\land Q)P\leq PQP$.  And $PQP\leq\uo(P,Q)$
follows from: $\uo(P,Q)-PQP=\uo(P,Q) -P\uo(P,Q)P=\uo(P,Q)(I-P)\geq 0$
recalling that $[\uo(P,Q),P]=0$.  

Eq.~(\ref{eq:1}) can be deduced from (\ref{eq:20}, \ref{eq:22}). The
latter also implies a version of sub- and super-additivity for
$\uo(P,Q)$ and $\lo(P,Q)$, respectively: 
\be
\no{subo}
{\sum}_a\,\uo(P_a,Q)\geq Q, \qquad
{\sum}_a\,\lo(P_a,Q)\leq Q, 
\ee
where $\sum_a P_a=I$. Hence $\lo(P,Q)$ and $\uo(P,Q)$ do not lead to
additive probabilities: the additive marginalization can still be
applied, but it leads to an upper bound $\sum_a\uo(P_a,Q)$ (and lower
bound $\sum_a\lo(P_a,Q)$) for the correct marginal probability
$\uo(I,Q)=\lo(I,Q)$; see (\ref{subo}). The correct marginals are
calculated from (\ref{eq:3}) by taking $P=I$ or $Q=I$. Note that
monotonicity does not hold, i.e.  generally
$\uo(P,Q)\not\leq\uo(I,Q)=Q$, though $P\leq I$. 

Eq.~(\ref{eq:22}) directly leads to (\ref{eq:4}, \ref{eq:5}).
The following feature is seen from (\ref{eq:20})
\be
\no{uni}
U\omega(P,Q)U^\dagger=\omega(UPU^\dagger,UQU^\dagger), \quad 
\omega=\lo,\uo,
\ee
where $U$ is a unitary operator: $UU^\dagger=I$. For further features of
the imprecise probability see \cite{armen,arshag}. Ref.~\cite{arshag}
shows that it is also consistent with the quantum conditional (two-time)
probability.  We emphasize that (\ref{eq:20}) are respectively the
minimal and maximal operators holding 
(\ref{borno}--\ref{eq:5}). If some of those conditions are omitted, the
imprecise probabilities can only become more precise.

{\bf Imprecise probability for combined systems}. We now turn to
applying the above formalism for two particles denoted by indices $1$
and $2$, living resp. in Hilbert spaces ${\cal H}_1$ and ${\cal H}_2$.
Let $P_k$ and $Q_k$ be two generally non-commuting ($[P_k,Q_k]\not=0$)
projectors living in ${\cal H}_k$, $k=1,2$. This is the standard set-up
for designing and studying Bell's inequalities \cite{fine}. For two
(non-identical) particle these projectors read
\be
\no{ord}
P_1\ot I,\qquad Q_1\ot I,\qquad I\ot P_2, \qquad I\ot Q_2. 
\ee
{Projectors refering to different particles naturally
commute, e.g.  $[P_1\ot I,I\ot P_2]=0$. Hence the joint feature is
described by the projector $(P_1\ot I)(I\ot P_2)=P_1\ot P_2$, which|being a
tensor product of single-particle operators|can be measured separately
in sub-systems (and then bringing the measurement data together). For
factorized states, $\rho_1\ot\rho_2$, these measurements lead to
uncorrelated probabilities. 

\comment{ We stress that all the operators in (\ref{ord}) are local,
since they refer to tensor products of operators from different spaces.
When averaged over factorized density matrices $\rho_1\otimes\rho_2$,
they will produce factorized (i.e. independent) probabilities. }

There are two ways (resp. (\ref{h1}) and (\ref{h2})) for
looking at the joint statistics of two local operators:
}
\begin{align}
\no{h1}
&\lo(P_1\ot P_2, Q_1\ot Q_2), & \uo(P_1\ot P_2, Q_1\ot Q_2); \\
\no{h2}
&\lo(P_1\ot Q_2, Q_1\ot P_2), & \uo(P_1\ot Q_2, Q_1\ot P_2).
\end{align}
For the lower probability operators in (\ref{h1}, \ref{h2}), 
(\ref{nogi}) leads to an intuitively expected outcome: 
\begin{align}
\no{h4}
&&\lo(P_1\ot P_2, Q_1\ot Q_2)=\lo(P_1,Q_1)\ot \lo(P_2,Q_2)\\
&&=\lo(P_1\ot Q_2, Q_1\ot P_2).
\no{h5}
\end{align}
Eq.~(\ref{h4}) shows that the lower probability operator is tensor
product of two local projectors. Eq.~(\ref{h5}) confirms that two
different ways (\ref{h1}) and (\ref{h2}) of combining commuting
projectors leads to the same outcome. Hence the lower probability is
local. 

The situation with the upper probability is different. Our
main result (deduced in Appendix) is formulated simpler after
joint block-diagonalization of $P_k$ and $Q_k$ (for each $k=1,2$) via a 
suitable unitary $U_k$ [cf.~(\ref{uni})]. This is known as the CS-representation 
\cite{dix,halmos,hardegree}: 
\be
\no{hu1}
P_k={\rm dg}[\P_k^{[2m_k]},I^{[m_k(1)]},0^{[m_k(2)]},I^{[m_k(3)]},0^{[m_k(4)]}],\\
Q_k={\rm dg}[\Q_k^{[2m_k]},I^{[m_k(1)]},I^{[m_k(2)]},0^{[m_k(3)]},0^{[m_k(4)]}],
\no{hu2}
\ee
where ${\rm dg}[A,B,...]$ means block-diagonal matrix with block
$A,B,...$. Upper indices are integers that indicate the dimension of
each square block matrix, e.g. $I^{[m_k(2)]}$ means $m_k(2)\times
m_k(2)$ unity matrix. We omit these indices, whenever their implications
are clear from the context. According to (\ref{hu1}, \ref{hu2}), the
original Hilbert space ${\cal H}$ can be represented as a direct sum of
subspaces with dimensions $2m_k$, $m_k(1)$, $m_k(2)$, $m_k(3)$ and
$m_k(4)$. Some of them can be zero, i.e. the subspaces will be absent
from (\ref{hu1}, \ref{hu2}). 

We recall
a possible explicit representation for the non-commuting parts $\P_k$ and $\Q_k$ of 
(resp.) $P_k$ and $Q_k$ as $2m_k \times 2m_k$ matrices 
\cite{dix,halmos,hardegree}:
\be
\P_k=\left(\begin{array}{cc}
C_k^2 & C_kS_k \\
C_kS_k & S_k^2 \\
\end{array}\right), ~
\Q_k=\left(\begin{array}{cc}
I & 0 \\
0 & 0 \\
\end{array}\right), 
\label{glen}
\ee
where $C_k$ and $S_k$ are invertible, self-adjoint $m_k\times m_k$ matrices holding
\be
\no{buka}
C_k^2+S_k^2=I,\qquad [C_k,S_k]=0. 
\ee
The simplest example of (\ref{hu1}--\ref{buka}) are the projectors for
$x$ and $z$-components of two spin-$\frac{1}{2}$ particles. Here ${\rm
dim}{\cal H}_k=2$, only the subspace with $2m_k=2$ is present, and
$\P_k=\frac{1+\sigma_x}{2}$, $\Q_k=\frac{1+\sigma_z}{2}$, where
$\sigma_x$ and $\sigma_z$ are the Pauli matrices.  Hence
$C_k=S_k=\frac{1}{\sqrt{2}}$ reduce to numbers. 

The following relations are deduced from (\ref{glen}, \ref{buka})
\be
\no{karakhan}
\P_k \lor \Q_k = I, ~~ \P_k \land \Q_k = 0, ~~ {\rm tr}\P_k ={\rm tr} \Q_k = m_k.
\ee
Hence in all other subspaces besides
$\P_k\lor \Q_k=I$ [cf.~(\ref{hu1})], the projectors $P_k$ and $Q_k$ commute. 

In the representation (\ref{hu1}, \ref{hu2}), the upper and lower probability operators
(\ref{eq:20}) read via (\ref{karakhan})
\begin{align}
\no{hu3}
\uo(P_k,Q_k)={\rm dg}(I-(\P_k-\Q_k)^2 ,I,0,0,0), \\
\no{hu33}
\lo(P_k,Q_k)={\rm dg}(0 ,I,0,0,0).
\end{align}

Here is our main result obtained within the CS representation
(\ref{hu1}, \ref{hu2}) (see Appendix):
\begin{eqnarray}
\label{lat}
&&\uo(P_1,Q_1)\ot \uo(P_2, Q_2)-\uo(P_1\ot P_2, Q_1\ot Q_2)~~~
\\
&&={\rm dg}\left(\, \uo(\P_1^\perp\ot \Q_2^\perp,\Q_1^\perp\ot \P_2^\perp),\,0,...,0\,\right)\geq 0,
\no{gajl}
\end{eqnarray}
where other 24 blocks nullify, and the inequality follows from the 
definition (\ref{eq:20}) of $\uo$.

Eq.~(\ref{gajl}) shows that the upper probability operator $\uo(P_1\ot
P_2, Q_1\ot Q_2)$ does not reduce to the tensor product $\uo(P_1,Q_1)\ot
\uo(P_2, Q_2)$, i.e. it does not have the local form. We take this as
indications of non-locality. This feature is stronger than the notion of
entanglement, since it is formulated without density matrices, i.e. on
the level of probability operators. Now $\uo(P_1\ot P_2, Q_1\ot Q_2)
=\uo(P_1,Q_1)\ot \uo(P_2, Q_2)$ is recovered whenever at least one pair
commutes, i.e. either $[P_1,Q_1]=0$ or $[P_2,Q_2]=0$ holds; e.g.
$[P_1,Q_1]=0$ implies that $\P_1$ and $\Q_1$ are absent from
(\ref{gajl}). Hence $\uo(P_1\ot P_2, Q_1\ot Q_2) \not=\uo(P_1,Q_1)\ot
\uo(P_2, Q_2)$ is due to non-commutativity at both $1$ and $2$. In
particular, we get (as we should) the correct marginal upper probability
$\uo(P_1,Q_1)\ot I$ for the particle $1$ whenever $P_2=Q_2=I$. 

\comment{The latter formula is the correct way (in contrast to
taking the additive sum) of marginalizing the imprecise joint
probability; recall the discussion around (\ref{subo}). }

{The inequality in (\ref{gajl}) means that the local form
$\uo(P_1,Q_1)\ot \uo(P_2, Q_2)$ of the upper probability is less precise
than the non-local expression $\uo(P_1\ot P_2, Q_1\ot Q_2)$, i.e.  upper
probabilities calculated via ${\rm tr}(\rho\,\uo(P_1,Q_1)\ot \uo(P_2,
Q_2)\,)$ will (for an arbitrary density matrix) be larger than those
calculated ${\rm tr}(\rho\, \uo(P_1\ot P_2, Q_1\ot Q_2)\,)$.  In this
sense, the imprecise probability reconciles locality with non-locality.
}

Eqs.~(\ref{gajl}) and (\ref{h1}, \ref{h2}) imply
\be
\no{kara}
\uo(P_1\ot Q_2, Q_1\ot P_2)-\uo(P_1\ot P_2, Q_1\ot Q_2)\not=0.
\comment{\nonumber\\
&&={\rm dg}[\, \uo(\P_1^\perp\ot \Q_2^\perp,\Q_1^\perp\ot \P_2^\perp)
\nn
&&-\uo(\P_1^\perp\ot \P_2^\perp,\Q_1^\perp\ot \Q_2^\perp)
,\,0,...,0\,]\not =0.
}
\ee
Hence depending on how we combine commuting projectors|via (\ref{h1})
or (\ref{h2})|we shall get different upper probability operators. This 
clearly contrasts with features (\ref{h4}, \ref{h5}) of the lower probability operator. As seen below,
(\ref{kara}) is generally non-zero even if we calculate the
upper probabilities (\ref{borno}) for independent subsystems, i.e. via
tensor-product states $\rho=\rho_1\ot\rho_2$.

\comment{Note that two measures of non-commutativity for each particle enter into
the right-hand-side of (\ref{kara}): $\p_k-\q_k$ ($\p_k'-\q_k'$) and
$[\p_k,\q_k]$ ($[\p_k',\q_k']$).}

{\bf Two spin-$1/2$ particles.} We now illustrate (\ref{gajl},
\ref{kara}) via a pair of spin-$1/2$ particles, where $m_k=1$ in
(\ref{hu1}, \ref{hu2}), and also $m_k(1)=m_k(2)=m_k(3)=m_k(4)=0$ for
$k=1,2$. For simplicity we also assume $P_1=P_2=P$ and $Q_1=Q_2=Q$ (the
index $k$ drops out), and $C_k=S_k=1/\sqrt{2}$ in (\ref{glen},
\ref{buka}), i.e. $P$ and $Q$ amount to $x$ and $z$ components of the
spin $\frac{1}{2}$. In (\ref{gajl}) we employ $P\lor Q=(P+Q)(P+Q)^-$
\cite{piziak}, where $X^-$ is the pseudo-inverse of matrix $X$
\cite{foo1}. We use definition (\ref{shun}) for the tensor product.
The result reads from (\ref{lat}) or from (\ref{gajl}):
\begin{align}
&\uo(P,Q)\ot \uo(P, Q)-\uo(P\ot P, Q\ot Q)\nonumber\\
\no{hh1}
&=\uo(P\ot Q, Q\ot P)=\frac{1}{12}\left(\begin{array}{cccc}
0 & 0 & 0 & 0 \\
0 & 2 & -1 & -1 \\
0 & -1 & 2 & -1 \\
0 & -1 & -1 & 2  \\
\end{array}\right),\\
&\uo(P,Q)\ot \uo(P, Q)-\uo(P\ot Q, Q\ot P)\nonumber\\
\no{hh2}
&=\uo(P\ot P, Q\ot Q)=\frac{1}{12}\left(\begin{array}{cccc}
1 & -1 & -1& 0 \\
-1 & 1 & 1 & 0 \\
-1 & 1 & 1 & 0 \\
0 & 0 &  0  & 3  \\
\end{array}\right).
\end{align}
Now both (\ref{hh1}) and (\ref{hh2}) have the same eigenvalues: $0$
and $1/4$ (both doubly degenerate).  However, they do
not commute. Let us define the following separable state:
\be
\rho\ot \rho, \qquad \rho=\left(\begin{array}{cc}
a & b\, e^{i\phi} \\
b \,e^{-i\phi} & 1-a \\
\end{array}\right),
\ee
where $\rho$ is a one-particle density matrix with real parameters $a$, $b$ and $\phi$ that hold
$1\geq a\geq 0$ and $a(1-a)\geq b^2$. These conditions ensure $\rho\geq 0$. 
The difference between (\ref{hh2}) and (\ref{hh1}) is not zero even for separable states:
\begin{align}
&{\rm tr}\left[\,(\uo(P\ot P, Q\ot Q)-\uo(P\ot Q, Q\ot P)\,)\,\rho\ot \rho\,\right]=\nonumber\\
&\frac{1}{12}\left[\,
 (1-2a)^2+4b^2+4b(1-2a)\cos(\phi)\,\right]\geq 0.
\label{brbr}
\end{align}
For separable states (\ref{brbr}) has a definite sign. No
definite sign is possible for all states, since ${\rm
tr}\left[\,\uo(P\ot P, Q\ot Q)-\uo(P\ot Q, Q\ot P)\,\right]=0$ due to the
fact that $\uo(P\ot P, Q\ot Q)$ and $\uo(P\ot Q, Q\ot P)$ have the same
eigenvalues.

{\bf Summary.} We studied the joint statistics of non-commuting
observables shared between two particles. This is a standard set-up for
defining quantum non-locality. Upon assuming the existence of joint
probabilities for non-commuting observables (or alternatively, specific
features of hidden variable theories), it leads to the Bell inequalities
for certain (entangled) states \cite{fine}. However, the precise joint
probability for non-commuting observables is denied by quantum mechanics
\cite{fine,arshag}, restricting the message of Bell inequalities to
inapplicability of certain non-quantum concepts to quantum mechanics. 

In contrast, we addressed the above set-up from the viewpoint of
imprecise joint probability for non-commuting observables. This concept
does exist: it is well-defined and holds all requirements asked by
quantum mechanics from a joint probability. Our basic message is that
for making more precise predictions for the joint probability of
two-particle observables requires non-locality, because the
corresponding upper probability is not a tensor product of one-particle
factors. In contrast to other forms of quantum non-locality (e.g. the
non-locality without entanglement \cite{nonloc}), the uncovered form of
non-locality is formulated on the level of observables, i.e.
independently from the notion of quantum states. Hence it survives for
separable states, as we saw for the simplest example.

{\bf Appendix.} For deriving (\ref{gajl}), we introduce two different types of tensor products:
\be
\no{len}
A\ot B\quad {\rm and}\quad 
A\oo B=\aoverbrace[9]{A\ot B}=B\ot A,
\ee
where $A\ot B$ is the usual from left-to-right tensor (Kroenecker) product,
e.g. when $A$ is a $2\times 2$ matrix, 
\be
\left(\begin{array}{cc}
a_{11} & a_{12} \\
a_{21} & a_{22} \\
\end{array}\right)\otimes B =
\left(\begin{array}{cc}
a_{11}B & a_{12}B \\
a_{21}B & a_{22}B \\
\end{array}\right),
\no{shun}
\ee
where $a_{ik}B$ (with $i,k=1,2$) means that a number $a_{ik}$ is
multiplied over all matrix elements of $B$. We stress that $A\ot B$ and
$A\oo B$ in (\ref{len}) are related by a unitary operator that does not
depend on $A$ and $B$. Now we note
\begin{align}
\no{go1}
& \aoverbrace[9]{\, {\rm dg}(A,B)\ot {\rm dg}(C,D)\,}= {\rm dg}(U,V) \times \\
\no{go2}
&{\rm dg}(A\ot {\rm dg}(C,D),\, B\ot {\rm dg}(C,D))\,{\rm dg}(U^\dagger,V^\dagger)  \\
\no{go3}
& =  {\rm dg}(A\oo {\rm dg}(C,D),\, B\oo {\rm dg}(C,D))  \\
\no{go4}
& = {\rm dg}(A\oo C, A\oo D, B\oo C, B\oo D),
\end{align}
where when going from (\ref{go1}) to (\ref{go2}) we used (\ref{shun}),
and where (\ref{go3}) is achieved from (\ref{go2}) by means of 
block-diagonal unitary matrices with blocks $U$ and $V$. 
Eq.~(\ref{go4}) is a block-diagonal matrix that will be useful below.

For $A$ and $B$ being projectors we get 
\be
\no{brutos}
&&{\rm dg}(A,B)\land {\rm dg}(C,D)={\rm dg}(A\land C,B\land D),\\
\no{go44}
&&{\rm dg}(A,B)\lor {\rm dg}(C,D)={\rm dg}(A\lor C,B\lor D),
\ee
where (\ref{brutos}) is derived from (\ref{nogi}), while (\ref{go44})
is deduced from (\ref{brutos}) via (\ref{gogi}).
Eqs.~(\ref{go1}--\ref{go44}) straightforwardly 
generalize to an arbitrary number of blocks.
Now recall from (\ref{uni}) that 
\begin{align}
&\aoverbrace[9]{\uo(P_1\ot P_2, Q_1\ot Q_2)}=
\uo(\aoverbrace[9]{P_1\ot P_2},\,\aoverbrace[9]{Q_1\ot Q_2}\,)\nn
&=
\aoverbrace[9]{P_1\ot P_2}\lor \aoverbrace[9]{Q_1\ot Q_2}-(\aoverbrace[9]{P_1\ot P_2}-\aoverbrace[9]{Q_1\ot Q_2})^2.
\no{hu10}
\end{align}
Using (\ref{go1}--\ref{go4}) and (\ref{hu1}--\ref{hu2}) we write down 
\be
\aoverbrace[9]{P_1\ot P_2}=&&{\rm dg}(\, \P_1\oo \P_2, \, \P_1, \,  0, \,  \P_1, \, 0,...)  \\
\aoverbrace[9]{Q_1\ot Q_2}=&&{\rm dg}(\, \Q_1\oo \Q_2,\, \Q_1,\,  \Q_1, \, 0, \, 0 , ...)
\ee
where for simplicity we write $\P_1\oo I$ as $\P_1$ and $I\oo I$ as $I$, and where other 20 blocks (in each
equation) were omitted for simplicity.  Now both $\aoverbrace[9]{P_1\ot
P_2}\lor \aoverbrace[9]{Q_1\ot Q_2}$ and $(\aoverbrace[9]{P_1\ot
P_2}-\aoverbrace[9]{Q_1\ot Q_2})^2$ are easy to calculate,
since they respect the block-diagonal structure; cf.~(\ref{go44}). Here 
we, in particular, employ $(\P_1\oo I)\lor (\Q_1\oo I)=I\oo I=I$; 
see the first equation in (\ref{karakhan}). Hence
\be
&&\aoverbrace[9]{P_1\ot P_2}\lor \aoverbrace[9]{Q_1\ot Q_2}=\nonumber\\
&&{\rm dg}(\P_1\oo \P_2\lor \Q_1\oo \Q_2,\,  I, \, \Q_1,\, \P_1,\, 0,  ...),\\
&&(\aoverbrace[9]{P_1\ot P_2}- \aoverbrace[9]{Q_1\ot Q_2})^2=\\
&&{\rm dg}(\,(\P_1\oo \P_2- \Q_1\oo \Q_2)^2,\,  (\P_1-\Q_1)^2, \, \Q_1,\, \P_1,\, 0,...)\nonumber
\ee
Employing (\ref{hu3}) we work out in the same way 
\be
&& \aoverbrace[9]{\uo(P_1, Q_1)\ot \uo(P_2, Q_2) }=\nonumber\\
&& {\rm dg}\left(\, (I-(\P_1-\Q_1)^2)\oo(I-(\P_2-\Q_2)^2),\right. \nonumber\\
&& \left.I-(\P_1-\Q_1)^2,\, 0,\, 0,\, 0,...\right).
\ee
These intermediate formulas lead finally to 
\begin{align}
&\aoverbrace[9]{\uo(P_1\ot P_2, Q_1\ot Q_2)}-
\aoverbrace[9]{\uo(P_1, Q_1)\ot \uo(P_2, Q_2) }\nn
&={\rm dg}(\,(\P_1\oo \P_2)\lor (\Q_1\oo \Q_2)- (\P_1\oo \P_2-\Q_1\oo \Q_2)^2 \nn
&-(I-(\P_1-\Q_1)^2)\oo(I-(\P_2-\Q_2)^2),\, 0,...,\, 0 \,),
\no{krot}
\end{align}
where the last 24 blocks are zero. Once only one block is non-zero, we obtain
\begin{align}
&\uo(P_1\ot P_2, Q_1\ot Q_2)-
\uo(P_1, Q_1)\ot \uo(P_2, Q_2) \nn
&={\rm dg}(\,(\P_1\ot \P_2)\lor (\Q_1\ot \Q_2)
- (\P_1\ot \P_2-\Q_1\ot \Q_2)^2 \nn
&-(I-(\P_1-\Q_1)^2)\ot(I-(\P_2-\Q_2)^2),\, 0,...,\, 0 \,).
\no{krotik}
\end{align}
Now (\ref{krotik}) can be simplified by employing 
\begin{align}
& (\P_1^\perp\ot \Q_2^\perp-\Q_1^\perp\ot \P_2^\perp)^2=
I- (\P_1\ot \P_2-\Q_1\ot \Q_2)^2 \nn
&-(I-(\P_1-\Q_1)^2)\ot(I-(\P_2-\Q_2)^2),
\label{kandagar}
\end{align}
whose derivation is algebraically tedious, but straightforward.
Below we shall prove that
\begin{gather}
\no{pushtu}
(\P_1\ot \P_2)\lor (\Q_1\ot \Q_2)+
(\P_1^\perp\ot \Q_2^\perp)\lor (\Q_1^\perp\ot \P_2^\perp)=I.
\end{gather}
Using (\ref{kandagar}, \ref{pushtu}) in (\ref{krotik}) finishes the proof of (\ref{gajl}). 

The proof of (\ref{pushtu}) is consists of three steps. First, we note
that the two projectors in left-hand-side of (\ref{pushtu}) are
orthogonal to each other thanks to
\be
\no{p1}
(\P_1\ot \P_2)(\P_1^\perp\ot \Q_2^\perp)=
(\Q_1\ot \Q_2)(\P_1^\perp\ot \Q_2^\perp)=0, \\
\no{p2}
(\P_1\ot \P_2)(\Q_1^\perp\ot \P_2^\perp) =
(\Q_1\ot \Q_2)(\Q_1^\perp\ot \P_2^\perp) =0.
\ee
Hence the left-hand-side of (\ref{pushtu}) is $\leq I$.
Second, we get using (\ref{nogi}, \ref{karakhan}) 
\begin{align}
\no{p3}
&(\P_1\ot \P_2)\land (\Q_1\ot \Q_2)=(\P_1\land \Q_1)\ot (\P_2\land \Q_2)\nonumber\\
&=(\P_1^\perp\ot \Q_2^\perp)\land (\Q_1^\perp\ot \P_2^\perp)=0.
\end{align}
Third, recall that if ${\cal S}_P$ is the Hilbert space generated by a
projector $P$, then ${\rm dim}[{\cal S}_P]={\rm tr} P$. Now aply to 
both sides of (\ref{pushtu}) the known formula
\cite{hardegree}
\begin{eqnarray}
  \label{eq:106}
  {\rm tr}(P\lor Q)+   {\rm tr}(P\land Q)=  {\rm tr}(P)+  {\rm tr}(Q),
\end{eqnarray}
that holds for any projectors $[P,Q]\not=0$, and use (\ref{p1}--\ref{p3}):
\begin{align}
\label{gusi}
&{\rm tr} \left(\, \P_1\ot \P_2\right)={\rm tr}\left(\, \Q_1\ot \Q_2\right)=m_1m_2, \\
&{\rm tr} \left(\,(\P_1\ot \P_2)\lor (\Q_1\ot \Q_2)\right)\nonumber\\
=&{\rm tr} \left(\,(\P_1^\perp\ot \Q_2^\perp)\lor (\Q_1^\perp\ot \P_2^\perp)\right)=2m_1m_2.
\label{gaga}
\end{align}
Eqs.~(\ref{gusi}, \ref{gaga}) show that the traces of both sides (=projectors) of (\ref{pushtu})
are equal. Hence (\ref{pushtu}) was proved.


\begin{thebibliography}{99}
\bibitem{foo1} If $X=UDU^\dagger$, where $U$ is unitary and $D$ is diagonal with elemenents $(d_1,..,d_n,0,...,0)$,
where $(d_1,..,d_n)$ are strictly non-zero, then $X^-=UD^-U^\dagger$, where $D^-$ is a diagonal matrix with elements 
$(1/d_1,..,1/d_n,0,...,0)$.
\end{thebibliography}
\end{document}